\documentstyle[12pt]{article}

\def\be{\begin{equation}}
\def\ee{\end{equation}}
\def\ba{\begin{array}{c}}
\def\ea{\end{array}}

\def\ben{$$}
 \def\een{$$}

\begin{document}

\titlepage

\begin{center}{\Large \bf
  ${\cal PT}-$symetrically regularized
Eckart, P\"{o}schl-Teller and Hulth\'{e}n potentials
 }\end{center}

\vspace{5mm}

\begin{center}
Miloslav Znojil
\vspace{3mm}

\'{U}stav jadern\'e fyziky AV \v{C}R, 250 68 \v{R}e\v{z},
Czech Republic\\

homepage: http://gemma.ujf.cas.cz/\~\,znojil

e-mail: znojil@ujf.cas.cz

\end{center}

\vspace{5mm}

\section*{Abstract}

${\cal PT}$ (= parity times time-reversal) symmetry of complex
Hamiltonians with real spectra is usually interpreted as a weaker
mathematical substitute for Hermiticity. Perhaps an equally
important role is played by the related strengthened analyticity
assumptions. In a constructive illustration we complexify a few
potentials solvable only in $s-$wave. Then we continue their
domain from semi-axis to the whole axis and get the new exactly
solvable models. Their energies come out real as expected. The new
one-dimensional spectra themselves differ quite significantly from
their $s-$wave predecessors.

\vspace{9mm}

\noindent
 PACS 03.65.Ge,
03.65.Fd

\vspace{9mm}


\newpage

\section{Introduction}

Our mathematical understanding of many physical systems can become
drastically simplified after their suitable complexification. This
is true, first of all, in the study of resonances \cite{Horacek}
and of several other quantum scattering phenomena \cite{Newton}.
Recently, the idea of working in a complexified phase space for
bound states \cite{BT} re-entered the scene with a new enthusiasm
supported by an immediate relevance of the related break-down of
parity ${\cal P}$ in certain field theories \cite{Milton}.

In the mathematically more accessible quantum mechanical models
certain exceptional complex interactions with ${\cal PT}$ symmetry
happen to become strictly equivalent to a real potential after a
supersymmetric \cite{Andrianov} or integral, Fourier-like
\cite{BG} transformation. For other models, the analysis of the
related purely real spectra of energies has been performed by
several techniques. One may recollect, e.g., the most
straightforward numerical experiments \cite{Alvarez},
semiclassical approximants \cite{BB} and the so called delta
expansions \cite{Bender}. Resummations of divergent perturbation
series \cite{Caliceti} and the so called exact WKB method
\cite{Pham} also offered several Hamiltonians for which the
spectra of energies $E_n$ were proved strictly real.

One of the most immediate sources of information about the
possible connection or correlation between the absence of a decay
Im $E_n= 0$ and the ${\cal PT}$ symmetry $H={\cal PT}H{\cal PT}$
itself is provided by the exactly solvable models in one
dimension. Step by step, there were proposed the ${\cal PT}$
symmetric versions of the harmonic oscillator \cite{BB}, of the
asymmetric Morse interaction \cite{Morse}, of the asymptotically
symmetric (sometimes called ``scarf") hyperbolic oscillator
\cite{Bagchi} and of its asymptotically asymmetric but locally not
too dissimilar (also known as Rosen-Morse) alternative
\cite{shapin}. Before complexification, all of them belong among
the so called shape invariant potentials (cf. the review
\cite{Khare}) so that some of  their properties can be clarified
using the language of supersymmetry \cite{Cannata}.

On the basis of numerical experience \cite{Bender} the current
attention is exclusively paid to the forces $V(x)$ which are
analytic in $x$. An extremely interesting byproduct of this point
of view can be found in a transition to more dimensions for
quartic (i.e., unsolvable) oscillators \cite{BG} and for the
central and exactly solvable ${\cal PT}$ symmetrized harmonic
oscillator \cite{PTHO} and Coulomb problem \cite{LevaiZ}. Within
the set of the similar forces with a centrifugal-like singularity
there still exist a few models without a clear interpretation.
After a glimpse in the Table 4.1 of the review \cite{Khare} we
immediately discover two of them, namely, the Eckart model
 \be
 V^{(Eck)}(r) = \frac{A(A-1)}{\sinh^2 r}-2B\frac{\cosh
x}{\sinh r}
 \label{Eckartp}
 \ee
and the generalized P\"{o}schl-Teller potential
 \be
V^{(PT)}(r) = -\frac{A(A+1)}{\cosh^2 r} +\frac{B(B-1)}{\sinh^2 r}.
\label{sPTP}
  \ee
In the standard interpretation \cite{Newton}, both these $s-$wave
models are only partially, {\em incompletely} solvable and, in
this sense, lie somewhere in a ``territory of nobody". This was
the main source of our present inspiration. We see no reason why
these two interactions should not be appropriately continued to
the whole line and classified, afterwards, as the two new or
``forgotten" exactly solvable ${\cal PT}$ symmetric models. In
detail, this will be done in sections \ref{sEcka} and \ref{ssPTP}
below.

We have to remind the reader that the later force (\ref{sPTP}) may
be often found in the current literature in its alternative form
$V^{(GPT)}(r) = {(u + v\,\cosh\,2\, r)}/{\sinh^2 2\, r}$
\cite{Khare} or in the special form known as Hulth\'{e}n potential
\cite{Newton}. The former correspondence is mediated by the
trivial re-scaling of the axis of coordinates by factor 2. In the
present context, the latter, much less trivial relationship
deserves a more explicit attention. Its thorough discussion will
be added here, therefore, in section \ref{shulth}. A few further
relevant overall comments  may be found in our summary and final
discussion in section 5.

\section{${\cal PT}-$regularization and Eckart oscillator
\label{sEcka} }

From the purely historical point of view the loss of Hermiticity
in the domain of complex couplings proved more than compensated by
the new insight in the solutions of one of the most popular
unsolvable models $V(x) = \omega\, x^2 + \lambda\,x^4$ \cite{Wu}.
Today, its spectrum is understood as a {\em single} multi-sheeted
analytic function of the complex coupling constant $\lambda \in
l\!\!\! C$. The same idea applies to the set of resonances in the
cubic well $V(x) = \omega\, x^2 + \lambda\,x^3$ although a careful
analytic continuation must be also performed in the coordinate $x$
itself \cite{Alvarez}. These observations guided the
semi-classical and numerical studies of the forces
$V^{(\delta)}(x) = \omega\,x^2+ g\,x^2 (i\,x)^{\delta}$ containing
a variable real exponent $\delta$ \cite{Bender}. The related
${\cal PT}-$symmetric quantum mechanics with its new perturbation
series \cite{pert} as well as quasi-classical approximation
schemes \cite{Pham} and matrix-truncation methods \cite{Guardiola}
works with the globally, asymptotically deformed paths of
integration in the related Schr\"{o}dinger equation
 \be
\left [-\,\frac{d^2}{dx^2} + V(x) \right ]
 \, \psi(x) = E  \, \psi(x).
\label{SE}
  \ee
The $\delta=2$, quartic anharmonic oscillator of ref. \cite{BBdva}
exemplifies these systems which need not remain integrable on the
real line. Its asymptotic integrability and decrease of wave
functions is only recovered after we bend both our coordinate
semi-axes downwards and replace
 \ben \{x \gg 1 \} \
\ \longrightarrow \ \ \{ x = \varrho\,e^{-i\,\varphi}\},
 \ \ \ \ \ \ \ \ \ \ \ \ \{x \ll -1
  \} \ \ \longrightarrow \ \ \{ x = - \varrho\,e^{i\,\varphi}\}
  \een
beyond certain distance $\varrho_0 \gg 1$ and within certain
bounds upon $\varphi \in (0, \pi/3)$. The further growth of
$\delta $ beyond $\delta =2$ would make both the asymptotical
$\varphi- $wedges shrink and rotate downwards in the complex
plane.

Let us defer the discussion of the similar cases to our last
section \ref{shulth} below. Returning now just to our first two
examples (\ref{Eckartp}) and (\ref{sPTP}) we may notice that both
of them may be characterized by a ``weak", $\varphi =0$ option.
Globally they do not leave the real axis of $x$ at all. Such a
simplification proves most natural in the $\delta \to \infty$
regular model of ref. \cite{mytri}), admitting the most natural
physical interpretation of the real physical coordinates after
all. The related ${\cal PT}-$symmetrized oscillators need not
necessarily differ from their Hermitian counterparts too much. One
can hope to encounter just slight modifications of the formulae
available, e.g., in the factorization context \cite{Infeld} and in
its Lie-algebraic \cite{Lie}, operator \cite{Dambrowska} or
supersymmetric \cite{Levai} re-interpretations.

Equally straightforward innovations may be expected in the domain
of our singular forces (\ref{Eckartp}) and (\ref{sPTP}). One can
simply {\em avoid} their isolated singularities by a {\em local}
deformation of the integration path. In this way the strong
repulsion in the origin (so popular in some phenomenological
models \cite{Hall} {\em and} fully impenetrable in one dimension)
becomes readily tractable via a suitable choice of the cut.

\subsection{Terminating solutions revisited}

Once we pay attention to the real $s-$wave potential
(\ref{Eckartp}) with the strongly singular core, usually
attributed to Eckart \cite{Eckart}, we have to keep in mind that
this Hermitian model is solvable on the half-line only, with $r
\in (0,\infty)$ and, conventionally, $A > 1/2$ and $B>A^2$. Its
fixed value of the angular momentum $\ell=0$ is in effect a
non-locality which lowers its practical relevance in three and
more dimensions.

As already mentioned, the local deformation of the integration
path enables us to forget about the strong singularity in the
origin. We may admit the presence of the so called irregular
components in $\psi(r)\sim r^{1-A}$ near $r=0$. They would be, of
course, unphysical in the usual formalism \cite{condit}. Here, on
the contrary, we continue $r \to x$ with $x \in (-\infty,\infty)$
and encounter the new possibilities.

In the new perspective we have to re-analyze the whole
Schr\"{o}dinger equation anew. Our choice of appropriate variables
 \ben
  \psi(x) = (y-1)^u (y+1)^v \varphi \left (\frac{1-y}{2} \right ),
  \ \ \ \ \ \ \ \ \
y=\frac{\cosh x}{\sinh x}=1-2z
  \een
is dictated by the arguments of L\'evai \cite{Levai}, and the
consequent ${\cal PT}-$symmetry considerations require that we use
the purely imaginary couplings $B = i \beta$. Then we insert
$V^{(Eck)}(x) $ in eq. (\ref{SE}) and our change of variables
leads to its new form
\be
z(1-z)\,\varphi''(z) +[c-(a+b+1)z]\,\varphi'(z) -ab\,\varphi(z)=0
 \ee
where
\be
c=1+2u, \ \ \ \ a+b=2u+2v+1, \ \ \ \ ab=(u+v)(u+v+1)+A(1-A)
  \ee
and
\be
4v^2=2B-E, \ \ \ \ \  \ \ \ 4u^2=-2B-E.
  \ee
Our differential equation is of the Gauss hypergeometric type and
its general solution is well known \cite{Ryzhik},
 \be
\varphi(z)= C_1\cdot\ _2F_1(a,b;c;z) + C_2\cdot\ z^{1-c}\
 _2F_1(a+1-c,b+1-c;2-c;z).
 \label{solu}
  \ee
Besides the obvious relevance of such an exceptional solvability
of a model with a strong singularity in quantum mechanics, an
independent encouragement of its study is provided by its
methodical appeal in the context of field theory, especially in
connection with the so called Klauder phenomenon \cite{Klauder}.

\subsection{Asymptotic boundary conditions}

Technically, the first thing we notice is that our parameters $a$
and $b$ are merely functions of the sum $u+v$ and vice versa,
$u+v=(a+b-1)/2$. The immediate insertion then gives the rule
$(a-b)^2=(2A-1)^2$ and we may eliminate
 \be
a = b \pm (2A-1).
 \label{lab}
 \ee
We assume that our solutions obey the standard oscillation
theorems \cite{Hille} and become compatible with the boundary
conditions $\psi(\pm \infty) =0$ in eq. (\ref{SE}) at a discrete
set of energies, i.e., if and only if the infinite series $_2F_1$
terminate. Due to the complete $a \leftrightarrow b$ symmetry, we
only have to distinguish between the two possible choices of
$C_2=0$  and $C_1=0$.

In the former case with the convenient $b = -N$ (= non-positive
integer) the resulting numbers $a+b$ and $u+v$ prove both real.
Using the definition of $B$ the difference $u-v=-i\beta/(u+v)$
comes out purely imaginary. The related terminating wave function
series (\ref{solu}), i.e.,
 \be
\psi(x) =
 \left ( \frac{1}{\sinh x} \right )^{u+v} \ e^{(v-u)x}
  \cdot\,
\varphi[z(x)]
 \label{soluc1}
  \ee
is asymptotically normalizable if and only if $u+v>0$. This
condition fixes the sign in eq. (\ref{lab}) and gives the explicit
values of all the necessary parameters,
 \be
 a=2A-N-1, \ \ \ \ u+v=A-N-1, \ \ \ \ u-v=-i\,\frac{\beta}{A-N-1}.
 \label{proto}
 \ee
For all the non-negative integers $N \leq N_{max}< A-1$ the
spectrum of energies is obtained in the following closed form,
 \be
 E = -\frac{1}{2}\,\left (u^2+v^2\right ) =
 -\left ( A-N-1 \right )^2 + \frac{\beta^2}{(A-N-1)^2},
 \ \ \ \ \ \ N = 0, 1, \ldots, N_{max}.
 \ee
The normalizable wave functions become proportional to Jacobi
polynomials,
 \be
\varphi[z(x)] = const.\cdot P^{(u/2,v/2)}_N(\coth x).
\label{polynom}
 \ee
We have shortly to return to the second option with $C_1=0$ in eq.
(\ref{solu}). Curiously enough, this does not bring us anything
new. Although the second Gauss series terminates at the different
$b=c-1 -N$, the factor $z^{1-c}$ changes the asymptotics and one
only reproduces the former solution. All the differences prove
purely formal. In the language of our formulae one just replaces
$u$ by $-u$ in (and only in) {\em both} equations (\ref{soluc1})
and (\ref{proto}). No change occurs in the polynomial
(\ref{polynom}).

\section{P\"{o}schl-Teller potential
 \label{ssPTP} }

Schr\"{o}dinger equation (\ref{SE}) with the bell-shaped potential
$V(r) \sim 1/{\cosh^2 r}$ belongs to the most popular exactly
solvable models in quantum mechanics. Its applications range from
the analyses of stability and quantization of solitons
\cite{Bullough} to phenomenological studies in atomic and
molecular physics \cite{physaa}, chemistry \cite{physbb},
biophysics \cite{physcc} and astrophysics \cite{physdd}.  Its
appeal involves the solvability by different methods \cite{Levai}
as well as a remarkable role in the scattering \cite{Newton}. Its
bound-state wave functions represented by Jacobi polynomials are
also encountered as super-partners of a complex ``scarf" model
\cite{Bagchi}.

Not too surprisingly, virtually all these applications lose their
physical ground after an addition of the repulsive spike. Still,
it is not too difficult to extend the exact solvability itself to
the latter potential called, often, the P\"{o}schl-Teller well
\cite{Poeschl}. The related Schr\"{o}dinger equation (\ref{SE})
must be confined to semi-axis $ r \in (0, \infty)$ or
appropriately regularized.

\subsection{Regularization}

We may repeat that the impossibility of using the real $V^{(PT)}$
of eq. (\ref{sPTP}) with $A > B > 0$ in more dimensions nor on the
whole axis in one dimension is felt unfortunate in methodical
considerations and in perturbation theory \cite{Harrell}.
Singularities of the centrifugal type are encountered in
phenomenological models \cite{Hall,Sotona} but, unfortunately, not
too many of them are solvable \cite{Mathieu}.

In our present regularization of the singularity we shall not
deform the straight integration path at all. We shall rather
proceed in a way inspired by the pioneering paper \cite{BG} where
Buslaev and Grecchi employed simply a constant downward shift of
the whole coordinate axis,
 \be
r = x-i\varepsilon, \ \ \ \ \ \  \ x \in (-\infty,\infty).
\label{shov}
  \ee
In a way similar to the oscillator $V^{(BB)}(x)=V^{(HO)}(x-ic) =
x^2 - 2icx - c^2$ of ref. \cite{BB} and to its three-dimensional
generalization \cite{PTHO} the meaning of the ${\cal PT}$ symmetry
degenerates here to the mere trivial invariance with respect to
the simultaneous reflection $x \to -x$ and complex conjugation $i
\to -i$. The shift (\ref{shov}) is the main source of a {\em
regularization} here. As long as $1/(x- i\varepsilon)^2 =
(x+i\varepsilon)^2 / (x^2+\varepsilon^2)^2 $ at any $\varepsilon
\neq 0$, the centrifugal term remains nicely bounded in a way
which is uniform with respect to $x$. Without any difficulties one
is able to work with the similar centrifugal-like terms on the
whole real line of $x$.

The same idea applies to the regularized P\"{o}schl-Teller
potential
 \ben V^{(RPT)}(x) = V^{(PT)}(x -
i\varepsilon), \ \ \ \ \ \ \ 0 < \varepsilon <\pi/2.
  \een
This potential is a simple function of the L\'{e}vai's
\cite{Levai} variable $g(r)=\cosh 2r$. As long as $g(x -
i\,\varepsilon) = \cosh 2x\,\cos 2\varepsilon - i\,\sinh 2x\,\sin
2\varepsilon$, the new force is ${\cal PT}$ symmetric on the real
line of $x \in (-\infty, \infty)$,
  \ben V^{(RPT)}(-x)= [V^{(RPT)}(x)]^*.
  \een
Due to the estimates $|\sinh^2(x-i\varepsilon)|^2 = \sinh^2 x
\cos^2 \varepsilon +\cosh^2 x \sin^2 \varepsilon = \sinh^2 x
+\sin^2 \varepsilon$ and $|\cosh^2(x-i\varepsilon)|^2 = \sinh^2 x
+\cos^2 \varepsilon$ the regularity of $V^{(RPT)}(x)$ is
guaranteed for any parameter $\varepsilon \in (0, \pi/2)$.

\subsection{Solutions}

In a way paralleling the the preceding section the mere analytic
continuation of the $s-$wave bound states does not give the
complete solution. One must return to the original differential
equation (\ref{SE}). There we may conveniently fix $A
+1/2=\alpha>0$ and $B-1/2= \beta>0$ and write
 \be
\left (-\,\frac{d^2}{dx^2} +
\frac{\beta^2-1/4}{\sinh^2 r(x)}
 -\frac{\alpha^2-1/4}{\cosh^2 r(x)}  \right )
 \, \psi(x) = E  \, \psi(x), \ \ \ \ \ \ r(x)= x-i\varepsilon.
\label{SEb}
 \ee
This is the Gauss differential equation
 \be
z(1+z)\,\varphi''(z) +[c+(a+b+1)z]\,\varphi'(z)
+ab\,\varphi(z)=0
\label{gauss}
 \ee
in the new variables
 \ben
\psi(x) = z^\mu(1+z)^\nu\varphi(z),
\ \ \ \ \ \ \ \ z = \sinh^2r(x)
 \een
using the suitable re-parameterizations
 \ben
\alpha^2=(2\nu-1/2)^2,
\ \ \ \ \ \ \ \ \
 \beta^2=(2\mu-1/2)^2,
\ \ \ \ \ \ \ \ \
 \een
 \ben
2\mu+1/2=c, \ \ \ \ \
2\mu+2\nu=a+b, \ \ \ \ \
E= -(a-b)^2.
 \een
In the new notation we have the wave functions
 \be
\psi(x) = \sinh^{\tau \beta+1/2}[r(x)]
 \cosh^{\sigma\alpha+1/2}[r(x)]\,\varphi[z(x)]
\label{formula}
  \ee
with the sign ambiguities $\tau = \pm 1$ and $\sigma=\pm 1$ in
$2\mu=\tau \beta+1/2$ and $2\nu=\sigma\alpha+1/2$. This formula
contains the general solution of hypergeometric eq. (\ref{gauss}),
 \be
\varphi(z) = C_1\ _2F_1(a,b;c;-z) + C_2z^{1-c}\
_2F_1(a+1-c,b+1-c;2-c;-z). \label{gensol}
  \ee
The solution obeys the complex version of the Sturm-Liouville
oscillation theorem \cite{Hille}.  In the case of the discrete
spectrum this means that we have to demand the termination of our
infinite hypergeometric series, suppressing its undesirable
asymptotic growth at $x\to\pm\infty$.

In a deeper analysis let us first put $C_2=0$. We may satisfy the
termination condition by the non-positive integer choice of
$b=-N$.  This implies that $a=N+1 +\sigma\alpha +\tau  \beta$ is
real and that our wave function may be made asymptotically
(exponentially) vanishing under certain conditions. Inspection of
the formula (\ref{formula}) recovers that the boundary condition
$\psi(\pm \infty) = 0$ will be satisfied if and only if
  \ben 1 \leq 2N+1\leq 2N_{max}+1<-\sigma
\alpha - \tau  \beta.
 \een The closed Jacobi polynomial
representation of the wave functions follows easily,
  \ben
\varphi[z(x)] =C_1\ \frac{N!\Gamma(1+\tau \beta)}{\Gamma(N+1+\tau
\beta)} \ P_N^{(\tau \beta,\sigma\alpha)}[\cosh 2r(x)].
  \een
The final insertions of parameters define the spectrum of
energies,
 \be
E=-( 2N+1+\sigma \alpha + \tau  \beta)^2 < 0. \label{energy}
 \ee
Now we have to return to eq.  (\ref{gensol}) once more. A careful
analysis of the other possibility $C_1=0$ does not recover
anything new.  The same solution is obtained, with $\tau$ replaced
by $-\tau$.  We may keep $C_2=0$ and mark the two independent
solutions by the sign $\tau$. Once we define the maximal integers
$N_{max}^{(\sigma,\tau)}$ which are compatible with the inequality
 \be
2N_{max}^{(\sigma,\tau)}+1< -\sigma\alpha-\tau \beta \label{maxes}
 \ee
we get the constraint $N \leq N_{max}^{(\sigma,\tau)}$. The set of
our main quantum numbers is finite.

\section{Bent contours and Hulth\'en potentials
\label{shulth}
}

In both our above examples (\ref{Eckartp}) and (\ref{sPTP}) an
overall ${\cal PT}$ symmetry of the Hamiltonian is, presumably,
responsible for the existence of the real and discrete spectrum
\cite{BB}. Cannata et al \cite{Cannata} and Bender et al
\cite{BBsqw} were probably the first to notice that one of the
various limits $\delta \to \infty$ of the power-law models with
$\varphi \to \pi/2-{\cal O}(1/\delta)$ becomes, unexpectedly,
exactly solvable again, in terms of special Bessel functions.
These observations attract attention to strongly deformed
contours. One of possibilities of their interpretation is the
Liouvillean change of variables \cite{Olver}.

\subsection{The ${\cal PT}$ symmetry preserving
changes of variables}

In the first step let us recollect that in the spirit of the old
Liouville's paper \cite{Liouville} the change of the (real)
coordinates (say, $r \leftrightarrow \xi$) in Schr\"{o}dinger
equation
 \be
\left[-\,\frac{d^2}{dr^2} + W(r)\right]\, \chi(r) = -\kappa^2
\,\chi(r)
 \label{SEor}
 \ee
mediates a transition to a different potential.  In terms of an
invertible function $r=r(\xi)$ which possesses a few first
derivatives $r'(\xi), \, r''(\xi), \ldots$ we get the new bound
state problem with the new interaction
\be
 V(\xi)-E=\left [
 r'(\xi)
 \right ]^2
 \left \{
 W[r(\xi)]+\kappa^2
 \right \}
+
 \frac{3}{4}
 \left [
 {r''(\xi) \over r'(\xi)}
 \right ]^2
 -
 \frac{1}{2}
 \left [
 {r'''(\xi) \over r'(\xi)}
 \right ]
 \label{newpot}
 \ee
and normalizable wave functions
 \be
\Psi(\xi)={ \chi[r(\xi)] \over \sqrt{r'(\xi)}}\ . \label{trenky}
 \ee
In the Jacobi-polynomial context the Liouvillean changes of
variables have been applied systematically to all the Hermitian
models (cf. Figure 5.1 in the review \cite{Khare} or ref.
\cite{Varshni} for a more detailed illustration).  A similar
exhaustive study is still missing for the ${\cal PT}$ symmetric
models within the same class. Let us now try to partially fill the
gap. For the sake of brevity we shall only restrict our attention
to the ${\cal PT}$ symmetric initial eq. (\ref{SEor}) with the
P\"{o}schl-Teller potential
 \be
 W(r)=\frac{\beta^2-1/4}{\sinh^2 r}
 -\frac{\alpha^2-1/4}{\cosh^2 r}, \ \ \ \ \ \ r = x -
 i\varepsilon,
 \ \ \ \ \ \ \ \ \ x \in (-\infty, \infty).
\label{PTex}
  \ee
Its normalizable bound states are proportional to the Jacobi
polynomials,
 \ben
\chi(r) = \sinh^{\tau \beta+1/2}r
 \cosh^{\sigma\alpha+1/2}r\,\
  P_n^{(\tau \beta,\sigma\alpha)}(\cosh 2r)
  \een
at all the negative energies $-\kappa^2 < 0$ such that
 \ben
 \kappa=
 \kappa^{(\sigma,\tau)}_n=-\sigma\alpha-\tau\beta -2n-1>0.
 \een
These bound states are numbered by $n = 0,1,\ldots, n_{max}^{
(\sigma,\tau)}$ and by the generalized parities $\sigma=\pm 1$
and $ \tau =\pm 1$.

We may note that our initial ${\cal PT}$ symmetric model
(\ref{SEor}) remains manifestly regular provided only that its
constant downward shift of the coordinates $r = r_{(x)} = x -
i\,\varepsilon$ remains constrained to a finite interval,
$\varepsilon \in (0,\pi/2)$. In a key step of its present
modification let us now change the coordinates as follows,
 \be
 \sinh r_{(x)} (\xi)= - i e^{i\xi}, \ \ \ \ \ \ \ \ \xi = v - iu.
\label{tren}
 \ee
This shifts and removes the singularity at $r=0$ to infinity ($u
\to +\infty$). In an opposite direction, one cannot proceed
equally easily from a choice of a realistic $V(\xi)$ to the
re-constructed coordinate $r(\xi)$. This methodical asymmetry is
due to the definition (\ref{newpot}) containing the third
derivatives and, hence, too complicated to solve. Still we are
quite lucky with our purely trial and error choice of eq.
(\ref{tren}). Firstly, the real line of $x$ becomes mapped upon a
manifestly ${\cal PT}$ symmetric curve $\xi = v - iu$ in
accordance with the compact and invertible trigonometric rules
 \ben
 \ba
 \sinh x \cos \varepsilon = e^u\,\sin v,\\
 \cosh x \sin \varepsilon = e^u\,\cos v,
 \ea
 \een
i.e., in such a way that
 \ben
 \ba
 v =\arctan \left (
\frac{\tanh x}{\tan \varepsilon} \right )= v_{(x)} \in \left
(v_{(-\infty)}, v_{(\infty)}\right )\equiv \left
(-\frac{\pi}{2}+\varepsilon, \frac{\pi}{2}-\varepsilon \right ),\\
u = u_{(x)} = \frac{1}{2} \ln \left ( \sinh^2x+\sin^2\varepsilon
 \right ) .
 \ea
 \een
Our path of $\xi $ is a down-bent arch which starts in its left
imaginary minus infinity, ends in its right imaginary minus
infinity while its top lies at $x=v=0$ and $-u=-u_{(0)}= \ln
1/\sin \varepsilon >0$. The top may move towards the singularity
in a way mimicked by the diminishing shift $\varepsilon \to 0$.
Although the singularity originally occurred at the finite value
$r\to 0$, it has now been removed upwards, i.e., in the direction
of $-u \to +\infty$.

\subsection{Consequences}

The first consequence of our particular change of variables
(\ref{tren}) is that it does not change the asymptotics of the
wave functions. As long as $r'(\xi)= i\tanh r(\xi)$ the
transition from eq. (\ref{SEor}) to (\ref{SE}) introduces just
an inessential phase factor in $\Psi(\xi)$. This implies that
the normalizability (at a physical energy) as well as its
violations (off the discrete spectrum) are both in a one-to-one
correspondence.

The explicit relation between the old and new energies and
couplings is not too complicated. Patient computations reveal
its closed form. With a bit of luck, the solution proves
non-numerical.  The new form of the potential and of its binding
energies is derived by the mere insertion in eq. (\ref{newpot}),
 \be
 V(\xi)=
 \frac{A}{(1-e^{2i\xi})^2}+\frac{B}{1-e^{2i\xi}},
\ \ \ \ \ \  E=\kappa^2.
 \label{hulth}
 \ee
At the imaginary $\xi$ and vanishing $A=0$ this interaction
coincides with the Hulth\'en potential.

In the new formula for the energies one has to notice their
positivity. This is extremely interesting since the potential
itself is asymptotically vanishing at both ends of its
integration path.  One may immediately recollect that a similar
paradox has already been observed in a few other ${\cal PT}$
symmetric models with an asymptotic decrease of the potential to
minus infinity \cite{BBdva,Sesma}.

The exact solvability of our modified Hulth\'en potential is not
yet guaranteed at all. A critical point is that the new couplings
depend on the old energies and, hence, on the discrete quantum
numbers $n$, $\sigma$ and $\tau$ in principle.  This could induce
an undesirable state-dependence into our new potential. Vice
versa, the closed solvability of the constraint which forbids this
state-dependence will be equivalent to the solvability at last. A
removal of this obstacle means in effect a transfer of the
state-dependence (i.e., of the $n-$, $\sigma-$ and
$\tau-$dependence) in
 \ben A=A(\alpha) = 1 - \alpha^2, \ \ \ \ \ \
\ \ C \ (= A +B) =\kappa^2-\beta^2 \een from $C$ to $\beta$.  To
this end, employing the known explicit form of $\kappa$ we may
re-write
\be
C=C(\sigma,\tau,n)= (\sigma\alpha+2n+1) (\sigma\alpha+2n+1
+2\tau \beta).
\ee
This formula is linear in $\tau\beta$ and, hence, its inversion
is easy and defines the desirable state-dependent quantity
$\beta=\beta(\sigma,\tau,n)$ as an elementary function of the
constant $C$.  The new energy spectrum acquires the closed form
\be
E=E(\sigma,\tau,n)=A+B+\frac{1}{4}\, \left [
\sigma\alpha+2n+1-\frac{A+B}{\sigma\alpha+2n+1} \right ]^2. \ee
Our construction is complete.  The range of the quantum numbers
$n, \ \sigma$ and $\tau$ remains the same as above.

\section{Discussion}

\subsection{Spectrum of the ${\cal PT}$ symmetric Eckart model}

The new spectrum of energies seems phenomenologically appealing.
The separate $N-$th energy remains negative if and only if the
imaginary coupling stays sufficiently weak, $\beta^2 < (A-N-1)^4$.
Vice versa, the highest energies may become positive, with
$E=E(N_{max})$ growing extremely quickly whenever the value of the
coupling $A$ approaches its integer lower estimate $1+N_{max}$
from above. In this way, even a weak ${\cal PT}$ symmetric force
$V^{(Eck)}(x)$ is able to produce a high-lying normalizable
excitation. This feature does not seem connected to the presence
of the singularity as it closely parallels the similar phenomenon
observed for the ${\cal PT}$ symmetric Rosen-Morse oscillator
which remains regular in the origin \cite{shapin}. Also, in a way
resembling harmonic oscillators the distance of levels in our
model is safely bounded from below. Abbreviating
$D=A-N-1=A_{effective}>0$ its easy estimate
 \ben
 E_N-E_{N-1}=(2D+1) \left (1 + \frac{\beta^2}{D^2(D+1)^2} \right )
 > 1 \een
(useful, say, in perturbative considerations) may readily be
improved to $ E_N-E_{N-1}>\beta^2/D^2$ at small $D\ll 1$, to $
E_N-E_{N-1}>2D$ at large $D\gg 1$ and, in general, to an algebraic
precise estimate obtainable, say, via MAPLE \cite{Maple}.

Let us emphasize in the conclusion that the formulae we obtained
are completely different from the usual Hermitian $s-$wave results
as derived, say, by L\'evai \cite{Levai}. He had to start from the
regularity in the origin which implied an opposite sign in eq.
(\ref{lab}). This had to end up with the constraint $B> 0$.
Moreover, the size of $B$ limited the number of bound states.

In the present ${\cal PT}$ symmetric setting, a few paradoxes
emerge in this comparison. Some of them may be directly related to
the repulsive real core in our $V^{(Eck)}(x)$ with imaginary $B$.
Thus, one may notice that the {\em increase} of the real repulsion
{\em lowers} the $N-$th energy. In connection with that, the
number of levels {\em grows} with the increase of coupling $A$. In
effect, the new bound-state levels emerge as decreasing from the
positive infinity (!). At the same time, the presence of the
imaginary $B = i\beta$ shifts the whole spectrum upwards precisely
in the manner known from non-singular models.

\subsection{Paradoxes in the P\"{o}schl Teller case}

Let us now compare our final result (\ref{energy}) with the known
$\varepsilon=0$ formulae for $s$ waves \cite{Levai}. An additional
physical boundary condition must be imposed in the latter singular
limit.  This condition fixes the unique pair $\sigma = -1$ and
$\tau = +1$. Thus, the set of the $s-$wave energy levels $E_N$ is
not empty if and only if $\alpha- \beta> 1$. In contrast, all our
$\varepsilon > 0$ potentials acquire a uniform bound
$|V^{(RPT}(x)| < const < \infty$. Due to their regularity, no
additional constraint is needed. Our new spectrum
$E^{(\sigma,\tau)}_N$ becomes richer. For the sufficiently strong
couplings it proves composed of the three separate parts,
 \ben
E^{(-,-)}_N< 0, \ \ \ \ \ 0 \leq N \leq N_{max}^{(-,-)}, \ \ \ \ \
\ \alpha+ \beta > 1,
 \een
 \be
E^{(-,+)}_N<0, \ \ \ \ \ \ 0 \leq  N \leq N_{max}^{(-,+)}, \ \ \ \
\ \ \alpha> \beta + 1,
 \ee
 \ben
E^{(+,-)}_N<0, \ \ \ \ \ \ 0 \leq  N \leq N_{max}^{(+,-)},
 \ \ \ \ \ \
\beta > \alpha+1.
 \een
The former one is non-empty at $ A + B > 1$ (with our above
separate conventions $A > -1/2$ and $B > 1/2$). Concerning the
latter two alternative sets, they may exist either at  $A> B$ or
at $B > A+2$, respectively. We may summarize that in a parallel to
the ${\cal PT}$ symmetrized harmonic oscillator of ref.
\cite{PTHO} we have the $N_{max}^{(-,+)}+1$ quasi-odd or
``perturbed", analytically continued $s-$wave states (with a nodal
zero near the origin) complemented by certain additional
solutions.

In the first failure of a complete analogy the number
$N^{(-,-)}_{max}+1$ of our quasi-even states proves systematically
higher than $N^{(-,+)}_{max}+1$, especially at the larger
``repulsion" $ \beta \gg 1$. This is a certain paradox,
strengthened by the existence of another quasi-odd family which
behaves very non-perturbatively.  Its members (with the ground
state $\psi_0^{(+,-)}(x) =\cosh^{A+1} [r(x)]\sinh^{1-B} [r(x)]$
etc) do not seem to have any $s-$wave analogue. They are formed at
the prevalent repulsion $B>A+2$ which is even more
counter-intuitive. The exact solvability of our example enables us
to understand this apparent paradox clearly. In a way
characteristic for many ${\cal PT}$ symmetric systems some of the
states are bound by an antisymmetric imaginary well. A successful
description of its perturbative forms $V(x) = \omega x^2+i\lambda
\,x^3$ \cite{Alvarez,Caliceti} carries numerous analogies with the
real and symmetric $V(x) = \omega x^2+ \lambda\,x^4$. The similar
mechanism creates the states with $(\sigma,\tau)=(+,-)$ in the
present example.

A significant novelty of our new model $V^{(RPT)}(x)$ lies in the
dominance of its imaginary component {\em at the short distances},
$x \approx 0$. Indeed, we may expand our force to the first order
in the small $\varepsilon>0$. This gives the approximation
 \be
\frac{1}{\sinh^2(x-i\varepsilon)}=
\frac{\sinh^2(x+i\varepsilon)}{(\sinh^2x + \sin^2\varepsilon)^2}
=\frac{1}{\sinh^2x}+2i\varepsilon \frac{\cosh x}{\sinh^3 x} +{\cal
O}(\varepsilon^2).
 \label{sini}
 \ee
We see immediately the clear prevalence of the imaginary part at
the short distances, especially at all the negligible $A = {\cal
O}(\varepsilon^2)$.

An alternative approach to the above paradox may be mediated by a
sudden transition from the domain of a small $\varepsilon \approx
0$ to the opposite extreme with $\varepsilon \approx \pi/2$. This
is a shift which changes $\cosh x$ into $\sinh x$ and vice versa.
It intertwines the role of $\alpha$ and $ \beta$ as a strength of
the smooth attraction and of the singular repulsion, respectively.
The perturbative/non-perturbative interpretation of both our
quasi-odd subsets of states becomes mutually interchanged near
both the extremes of the parameter $\varepsilon$.

The dominant part (\ref{sini}) of our present model leaves its
asymptotics comparatively irrelevant. In contrast to many other
${\cal PT}$ symmetric models as available in the current
literature our potential vanishes asymptotically,
 \ben V^{(RPT)}(x) \to 0, \ \ \ \ \ \ \ \ x \to \pm \infty.
  \een
An introduction and analysis of continuous spectra in the ${\cal
PT}$ symmetric quantum mechanics seems rendered possible at
positive energies. This question will be left open here.

In the same spirit we may also touch the problem of the possible
breakdown of the ${\cal PT}$ symmetry. In our present solvable
example the violation of the ${\cal PT}$ symmetry is easily
mimicked by the complex choice of the couplings $\alpha$ and $
\beta$. Due to our closed formulae the energies will still stay
real, provided only that ${\rm Im}\ (\sigma\alpha+ \tau \beta)=0$.

\subsection{Transition to the Hulth\'{e}n model}

In the light of our new results we may now split the whole family
of the exactly solvable ${\cal PT}$ symmetric models which contain
a strong singularity in the two distinct categories.  The first
one ``lives" on the real line and may be represented or
illustrated not only by the popular Laguerre-solvable harmonic
oscillator \cite{PTHO} but also by both our present
Jacobi-solvable forces. The second category requires a arch-shaped
path of integration which lies confined within a narrow vertical
strip. It also involves both the Laguerre and Jacobi solvable
subsets. The former one may be represented by the complex Morse
model of ref. \cite{Morse} and by the Coulomb force with a complex
charge \cite{LevaiZ}. Our present new Hulth\'en example offers
their first Jacobi-solvable counterpart.  The parallels may be
illustrated by the following picture
 \ben \ba
\\
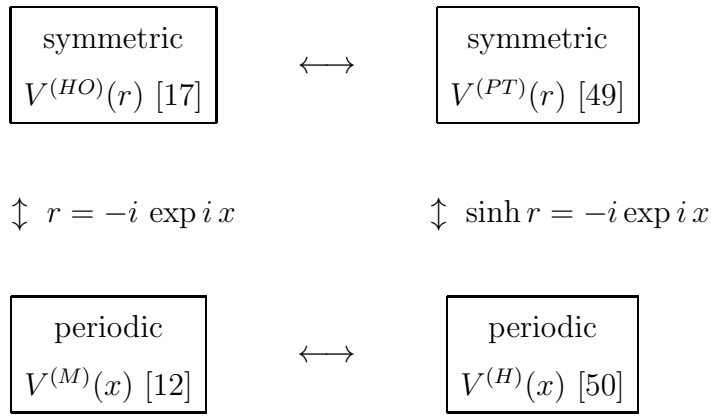
\begin{array}{|c|} \hline
{\rm symmetric}\\ V^{(HO)}(r) \ \cite{PTHO}\\ \hline \ea \ \ \ \ \
\ \ \stackrel{}{ \longleftrightarrow } \ \ \ \ \ \ \
\begin{array}{|c|} \hline {\rm symmetric}\\
V^{(PT)}(r)
 \ \cite{PTP}\\ \hline \ea
\\
\\ \ \ \ \  \  \ \ \ \
  \updownarrow \ r = -i\, \exp i\,x  {\rm }\
 \ \ \  \ \ \ \  \ \ \ \ \
\ \ \ \ \  \updownarrow \
 \sinh r= - i \exp {i\,x}
   {\rm }\  \ \\
\\
\begin{array}{|c|} \hline
{\rm periodic}\\ V^{(M)}(x)
 \
\cite{Morse}\\ \hline \ea \ \ \ \ \ \ \ \  \stackrel{}{
\longleftrightarrow } \ \ \ \ \ \ \ \ \begin{array}{|c|} \hline
{\rm periodic}\\ V^{(H)}(x)
\
\cite{hulth}\\ \hline \ea
\\
\\
\ea
 \een
where the vertical correspondence originates from the changes of
variables. One notices the similarities in the (symmetric or
periodic) form of the functions $V$ as well as the differences in
the straight-line or bent-curve shapes of the domains $r=r(t) \in
l\!\!\!C$ or $x=x(t) \in l\!\!\!C$, respectively.

The less formal difference between the two categories may be also
sought in their immediate physical relevance.  Applications of the
former class may be facilitated by a limiting transition which is
able to return them back on the usual real line.  In contrast, the
second category may rather find its most useful place in the
methodical considerations concerning, e.g., field theories and the
mechanisms of the parity breaking \cite{Milton}.  Within the
quantum mechanics itself the second category might also parallel
the studies of the ``smoothed" square wells in non-Hermitian
setting \cite{Cannata,BBsqw}.

In the conclusion let us recollect that the ${\cal PT}$ symmetry
of a Hamiltonian replaces and, in a way, {generalizes} its usual
Hermiticity. This is the main reason why there exists an
unexplored space for new solvable models.  In their context, an
example with an ``intermediate", hyperbola-shaped arc of
coordinates remains still to be discovered.  Up to now this type
of contour has only been encountered in the ``quasi-solvable"
(i.e., partially numerical) model of ref. \cite{BBdva}.

\section*{Acknowledgement}

Partially supported by the grant Nr. A 1048004 of the Grant
Agency of the Academy of Sciences of the Czech Republic.

\end{document}